\documentstyle[prd,aps,twocolumn,epsf]{revtex}
\def\br{\begin{eqnarray}}
\def\er{\end{eqnarray}}
\def\be{\begin{equation}}
\def\ee{\end{equation}}
\def\a{\alpha}
\def\D{\Delta}

\def\G{\Gamma}

\def\L{\Lambda}

\def\({\left(}
\def\){\right)}
\def\s{\sigma}

\begin{document}
\twocolumn[\hsize\textwidth\columnwidth\hsize\csname 
@twocolumnfalse\endcsname                            
%
%
\title{
Critical coupling for chiral symmetry breaking 
in QCD motivated models}
\author{
A.~C.~Aguilar, A.~A.~Natale and R.~Rosenfeld \\
}
\address{
Instituto de F\'{\i}sica Te\'orica\\ 
Universidade Estadual Paulista\\
Rua Pamplona 145\\ 
01405-900, S\~ao Paulo, SP\\
Brazil}
\date{\today}
\maketitle 
\begin{abstract}
We determine the critical coupling constant above which dynamical 
chiral symmetry breaking occurs in a class of QCD motivated
models where the gluon propagator has an enhanced infrared
behavior. Using methods of bifurcation theory we find that
the critical value of the coupling constant is always smaller
than the one obtained for QCD. 
\end{abstract}


\vskip 0.5cm]                           
\section{Introduction}

\noindent

The hadronic physics at low energy
is expected to be described
by the infrared properties of Quantum Chromodynamics (QCD), whose
non-perturbative character in general forces us to make use of 
approximate models in order to understand the strong interaction
effects at a small momentum scale. One of these models became
known as Global Color Model (CGM)~\cite{tandy} -\cite{cahill},
which is a quark-gluon quantum field
theory that describes QCD for low energy
processes.
There are many recent calculations exemplifying the 
remarkable success of this procedure~\cite{rob,fr}. It relates the
hadronic properties to the Schwinger functions of quarks and
gluons. Therefore, when comparing the theoretical calculations
to some low energy data, {\it e.g.} pseudoscalars
masses and decay constants or other chiral parameters, we 
are learning about the
infrared behavior of the quark and gluon propagators.
In the near future, this semi-phenomenological tool may 
reveal to be even more successful than the relativistic quark 
model or the bag model. 

In order to have an idea of what is behind the GCM, we can recall that its
action is obtained from the QCD generating
functional in the standard way~\cite{tandy}, with the main
difference being that in the functional generator of the
connected gluon $n$-point functions we neglect 
the higher than two $n$-point functions, expecting that a
phenomenological description of the gluon propagator $g^2 D_{\mu\nu}(x-y)$
retains most of the information about the non-Abelian character of QCD.
The effect of this approximation can only be measured in
comparison with experiments, and in fact it does work quite well once
we model appropriately the infrared behavior of the gluon propagator.

Some of the gluon propagators used in the GCM have a quite enhanced 
infrared behavior. One example is the introduction of a delta
function $\delta (k)$ as prescribed in Ref.~\cite{munczek},
which is a confining propagator according to the criterion
of absence of real $k^2$ poles for the 
quark propagators~\cite{munczek,pagels}. Another ansatz for the
gluon two-point function in the infrared is 
$g^2 D(k^2) = 3 \pi^2 (\chi/\Delta)^2 \exp{(-k^2/\Delta )}$ ~\cite{meissner},
which was inspired by and approaches the $\delta$ function ansatz of
Ref.~\cite{munczek}  for $\Delta \rightarrow 0 $, where $\chi$ and $\Delta$
are adjustable parameters. In detailed calculations of 
chiral parameters more complete ansatze for the
gluon propagators than the above ones have been used, in general including the
asymptotic behavior of the gluon propagator as predicted by QCD. In
Ref.~\cite{fr} the following form for the gluon propagator in the Landau gauge
was introduced 
\be
g^2D_{\mu\nu}(k) = \left\{ \delta_{\mu\nu} -\frac{k_\mu k_\nu}{k^2}\right\}
D (k^2) \; ,
\label{e1}
\ee
where,
\br
D(k^2) &\equiv & \frac{g^2}{k^2[1+\Pi (k^2)]}
\nonumber \\
&=&  4\pi^2 d\left[ 4\pi^2 m_t^2\delta^4(k) +
\frac{1-e^{(-k^2/[4m_t^2])}}{k^2}\right]\;,
\label{e2}
\er
with $d=12/(33-2n_f)$, and $n_f=3$ (considering only three quark flavors).  
The mass scale $m_t$ determined in Ref.~\cite{fr} was interpreted
as marking the transition between the perturbative and
nonperturbative domains. Another ansatz is~\cite{maris}
\be
g^2 D(k^2) = 3 \pi^2 \frac{\chi^2}{\Delta^2} \exp {(-k^2 )}
+ \frac{\alpha_s(k^2)}{k^2}  {\cal{F}}(k^2),
\label{e3}
\ee
where 
\be
\alpha_s (k^2) = \frac{4 \pi^2 d g^2}{ \ln {(\frac{k^2}{\Lambda^2} + \tau )}},
\label{rc}
\ee ${\cal{F}}(k^2)$
is a function chosen differently in the papers of
Ref.~\cite{maris}, $\alpha_s (k^2)$ describes the QCD running coupling
constant where $ \tau $
is a parameter adjusted phenomenologically. Notice that in the above
expressions the  momenta are in Euclidean space. Of course, there are still
other variations of these ansatze~\cite{silva} and attempts to explain the
enhanced  behavior in the infrared~\cite{pavel}.

In this work we study the bifurcation of the quark self-energy
within the context of the GCM, {\sl i.e.} we determine the critical coupling
constant of the truncated Schwinger-Dyson equation for the quark propagator
above which dynamical chiral symmetry breaking occurs using the gluon
propagators discussed in the previous paragraph. It is known from
analytical and numerical studies of the Schwinger-Dyson equations that
dynamical chiral symmetry breaking takes place in QCD when the coupling
constant ($\alpha_s$) is of 
$O(1)$~\cite{miransky,atkinson,roberts}. These studies have also been
performed with the use of effective potentials~\cite{peskin} and corroborated
by lattice simulations~\cite{kogut}. Therefore, it is natural to ask what is
the critical coupling in the GCM with the phenomenological gluon propagators
proposed in the literature. 

We will apply the standard techniques of
bifurcation theory as used by  Atkinson and
collaborators~\cite{atkinson,atkinson2,dif,agm} and verify that for the gluon
propagator given by a delta function the chiral symmetry is always broken no
matter what ``perturbative" propagator is added to the $\delta$. For a
propagator of the form given by Eq.(\ref{e3}) we obtain a lower bound
on the critical coupling smaller than the one obtained  for
QCD~\cite{atkinson2}, and evaluate the smallest characteristic number 
confirming the result obtained with the $\delta$ function propagator
in the limit that the gaussian width goes to zero. In
Section II we discuss  chiral symmetry breaking for an infrared gluon
propagator similar to Eq.(\ref{e2}) using a bifurcation analysis of the
Schwinger-Dyson equations. In Section III we analyse the infrared
gluon propagator given by a gaussian form using a different technique. The
last section contains our conclusions.

\section{The critical coupling for a confining propagator}       

\noindent

The Schwinger-Dyson equation for a massless quark propagator in Minkowski space
can be written in the form
\begin{equation}
S^{-1}(p) = \not\!{p} - \imath g^2 \int {d^4q\over (2
\pi)^4}D^{ab} _{ \mu \nu}(p-q) \Gamma^\mu(q,p){\lambda_c^a\over
2}S(q)\gamma ^\nu{\lambda_c^b\over 2}, 
\label{edse}
\end{equation}
where the gluon propagator in the Landau gauge, which will be used
throughout the paper, is given by
\begin{equation}
D_{\mu\nu} ^{ab}(k) = \delta^{ab}\left[{-g_{\mu\nu} +{ k_\mu k_\nu\over
k^2}}\right]{1\over k^2 +i\epsilon}. 
\label{gluon}
\end{equation}

In the case that the gluon propagator is given  by
the confining delta function~\cite{munczek}
\begin {equation}
 D_{\mu\nu}^{ab}(k)= \delta^{ab} \left[ g_{\mu\nu} - {k_\mu k _\nu \over
{k^2} }\right] \beta \delta^4(k) , 
\label{delta}
\end{equation}
where $\beta$ is an adjustable dimensional parameter. Note that
the term $k_\mu k_\nu\delta(k)/k^2$ is undefined from the point of view of
generalized functions. Of course, we could work with the Feynman gauge where
this unwanted behavior is softned. Fortunately the $\delta$ function gives an
integrable singularity and only because of this peculiarity we have not a
strong pathological behavior. With Eq.(\ref{delta})
it is quite easy to verify that the quark self-energy has a nontrivial
solution for any positive value of $g^2$, in the rainbow approximation
$(\Gamma^\mu(q,p) = \gamma ^\mu)$.
With the quark propagator given by 
\begin {equation}
{S} ^{-1} (p) = A(p^2)\not\!{p} - B(p^2),
\label{propag}
\end{equation} 
and the gluon propagator of Eq.(\ref{delta}) it follows from Eq.(\ref{edse}),
in Euclidean space, the set of coupled integral equations: 
\begin {eqnarray}
[A(p^2)-1]p^2  = {4\over3} g^2 \beta \int {d^4q\over (2
\pi)^4}\delta^4(p-q){A(q^2)\over q^2A^2(q^2) + B^2(q^2)} \nonumber \\
\times\bigg[p\cdot q + 2{q\cdot(p-q)(p-q)\cdot p\over(p-q)^2}\bigg] \nonumber\\
B(p^2) = 4 g^2 \beta \int {d^4q\over (2
\pi)^4}\delta^4(p-q){B(q^2)\over q^2A^2(q^2) + B^2(q^2)}.
\label{coupled}
\end{eqnarray} 

The solutions of the nonlinear coupled integral equations
were determined in
Ref.\cite{tandy,robwil}, showing that $ A(p^2)$ is a constant. Without loss of
generality we can set $ A(p^2) = 1 $ and  verify the behavior of
the self-energy apart from a normalization constant.
In this case we obtain the following equation for $B(p^2)$
\begin {equation}
B(p^2) = {4 g^2 \beta \over (2\pi)^4} \int
d^4q{B(q^2)\delta^4 (q-p) \over q^2 + B^2(q^2)}, 
 \label{integd}
\end{equation}
whose integration leads to
\begin {equation}
B(p^2) = {4 g^2 \beta \over (2\pi)^4} {B(p^2)\over p^2 +
B^2(p^2)}. 
\label{pol2}
\end{equation}

Equation (\ref{pol2}) has the solution
\[ B(p^2) = \left\{ \begin{array}
{r@{\quad:\quad}l}
 \sqrt{{4 g^2 \beta\over (2\pi)^4} - p^2 }  & p^2 \le {4 g^2
\beta\over (2\pi)^4} \\  0 & p^2 > {4 g^2 \beta\over(2\pi)^4},
\end{array} \right.  \]	
for any non-zero and positive value of the coupling constant. If
we consider only this infrared behavior of the gluon propagator
we would conclude that this model is inconsistent with previous
work on QCD, which predicts a critical value of the coupling constant for
the  onset of chiral symmetry breaking. However, the ultraviolet part
of the gluon propagator cannot be neglected and the use of the
following propagator is frequent~\cite{robwil,fr}
\begin {eqnarray}
 D_{\mu\nu}^{ab}(k)= \delta^{ab}\bigg[ \delta_{\mu\nu} - {k _\mu k_\nu \over
{k^2} }\bigg]\beta \delta^4(k)  \nonumber \\
 + \delta^{ab}\bigg[  \delta_{\mu\nu} - {k _\mu k
_\nu \over {k^2}}\bigg]{1\over{k^2}}. 
\label{deltapert}
\end{eqnarray}
The following comments are in order: a) In some papers the coupling
constant $g^2$ is absorbed in the parameter $\beta$~\cite{fr}.
Here we assume that $g^2$ can always be factorized, and the
critical behavior is going to be related to it. b) We
assume a constant $g^2$. In the next section it will be shown that
our result does not change with the inclusion of the running coupling
in the ultraviolet part of the gluon propagator.
c) This propagator has been used in the literature with some minor
differences; these variations are not important for our
result as will become clear in the next section. 

When the gluon propagator is given by Eq.(\ref{deltapert}) we need a 
thorough analysis of the quark self-energy to find for which value of the
coupling constant the quark self-energy  admits a nontrivial solution. 
The substitution of Eq.(\ref{deltapert}) into Eq.(\ref{edse}) yields
\begin {equation}
B(x) = { g^2 \over 4\pi^2}\bigg( {\kappa B(x)\over x +
B^2(x)} + \int_{\mu^2}^{\L^2} {dy\over
x_{max}}{yB(y)\over y +B^2(y)}\bigg), 
\label{completa1}
\end{equation}
where  $ x = p^2 $ , $ y = q^2 $, $ x_{max} = \max(x,y)$ and 
$ \kappa = {\beta\over \pi^2} $. We will be looking
for solutions of Eq.(\ref{completa1}) only in the interval $[\mu^2 , \L^2]$,
although the $\delta$ function has been integrated in the full
range of momenta. A rigorous handling of this integration does not change
the conclusions about the bifurcation point in the limit of very small (large)
infrared (ultraviolet) cutoff.

It is important to verify
that Eq.(\ref{completa1}) is a very peculiar one, in the sense that for very
small values of $g^2$, keeping the product $g^2 \kappa$ fixed, we do have the
solution of the nonlinear equation, which, at low momenta, approaches
\begin {equation}
B(x \rightarrow 0) = \( \sqrt{g^2 \kappa\over {4\pi^2}} \)
_{g^2 \rightarrow 0 \, , \, g^2 \kappa \ fixed }  .
\label{pb}
\end{equation}

To find nontrivial small solutions of the nonlinear equation
(\ref{completa1}) we study the linearized equation, {\sl i.e.}
the functional derivative of Eq.(\ref{completa1}) evaluated at
$B (x) = 0 $~\cite{atkinson,atkinson2,dif,agm}. Writing $ \delta B = f $, 
we obtain
\begin{equation}
f(x) = { g^2 \over 4\pi^2}\bigg( {\kappa f(x)\over x} +
\int_{\mu^2}^{\L^2} {dy {f(y)\over x_{max}}}\bigg).
\label{linear}
\end{equation}
Note that we are not considering the equation for 
$A(p^2)$ because this equation is of second order in $ \delta B$ ~\cite{agm}.
The existence of a solution for Eq.(\ref{linear}) is a sufficient condition
for the onset of chiral symmetry breaking~\cite{atkinson}.
Defining $ \alpha\equiv {g^2 / 4\pi^2 = \a_s/\pi} $,  
Eq.(\ref{linear}) is equivalent to the differential equation 
\begin{equation}
x(x - \alpha \kappa )f^{''}(x) + 2xf^{'}(x) + \alpha f(x)=0,
\label{diferencial}
\end{equation}
together with the
ultraviolet boundary condition $ [(x-\a\kappa)f(x)]'|_{x\rightarrow \L^2} =
0$ and another condition valid in the infrared region.

The choice of the infrared boundary condition is a crucial one. Note that
if we neglect the second term of the right-hand side of Eq.(13) we only 
obtain a trivial result, no matter what is the boundary condition
coming from Eq.(13). However, we know that the nonlinear equation always has a
nontrivial solution for the $\delta$ function propagator. The 
most suitable infrared  boundary condition is the one that comes from the
nonlinear integral equation (12), since when applied to the linear equation
the result must be consistent with the known solution of the nonlinear
equation for small $g^2$ (see Eq.(\ref{pb})). Therefore, our infrared
boundary condition is given by
\be
f(x)|_{x\rightarrow \mu^2} = \sqrt{\a\kappa} .
\label{iv}
\ee

Equation (\ref{diferencial}) can be put in the form of a hypergeometric
equation performing a shift $x \rightarrow z + \a
\kappa$ and defining $y = -z / \a \kappa$ leading to
\br
y(1-y)f^{''}(-\a\kappa y +\a\kappa) + 2(1-y)f^{'}(-\a\kappa y +\a\kappa)
\nonumber \\
 + \alpha f(-\a\kappa y +\a\kappa)=0, 
\label{hiperg}
\er

The solution of Eq.(\ref{hiperg}) that obeys the infrared boundary
condition is
\begin{equation}
f(x) =  A F \left(a,b;2;1-{x\over\alpha \kappa }\right),
\label{solfinal}
\end{equation}
where 
\be
A=\sqrt{\a\kappa}\G(3/2 -\s)\G(3/2 +\s),
\label{normal}
\ee
is a normalization constant, $\s$ is defined as
$ \sigma = ({1\over4} - \alpha)^{1/2} $, and
\begin{eqnarray}
a= {1\over 2} + \sqrt{{1\over 4} - \alpha},\nonumber \\
b= {1\over 2} - \sqrt{{1\over 4}- \alpha}. 
\label{constantes}
\end{eqnarray}
To apply the ultraviolet boundary condition we recall the
following relation involving hypergeometric functions
\be
{d\over dz}[z^{c-1}F(a,b;c;z)]=(c-1)z^{c-2}F(a,b;c-1;z),
\label{deriv}
\ee
which lead us to the analysis of the zeros of the following equation
\br
[(x&-&\a\kappa)f(x)]'|_{x\rightarrow \L^2} =  \nonumber \\ 
 &-&\a\kappa A F\left(1/2 + \s,1/2
-\s;1,1-{x\over\a\kappa}\right)\bigg|_{x\rightarrow \L^2}  . \label{uvc}
\er

We need to consider the solution of Eq.(\ref{uvc}) in three 
different regions of the parameter $ \alpha $, namely, $0 < \alpha <
{1\over4}$, $\alpha ={1\over4}$ and $\alpha >{1\over4}$, and
study the asymptotic behavior of $ f(x) $ in each one of
these cases. When $ \alpha < {1\over 4} $ the relation
\begin{eqnarray}
&&F(a,b;c;z)= \nonumber \\
&&{\Gamma(c)\Gamma(b-a)\over\Gamma(b)\Gamma(c-a)}(-z)^{-a}F(a,1+a-c;1+a-b;z^{-1}) \nonumber \\
&&+{\Gamma(c)\Gamma(a-b)\over\Gamma(a)\Gamma(c-b)}(-z)^{-b}F(1+b-c,b;1+b-a;z^{-1}),
\label{assintotica}
\end {eqnarray}
can be used together with 
\be
F(a,b;c;0) = 1 
\label{zero}
\ee
to give
\br
&[&(x-\a\kappa)f(x)]'|_{x\rightarrow \L^2}  = \nonumber \\
&&A'\bigg({x\over\alpha \kappa } - 1 \bigg)^{-1/2 +
\sigma} + B'\bigg({x\over \alpha \kappa } - 
1\bigg) ^{-1/2 - \sigma}\bigg|_{x\rightarrow \L^2}, 
\label{menor1/4}
\end{eqnarray}
where $ A'$ and $ B'$ depend on $\s$ and $\a\kappa$.

In this case $\s$
is a real and positive number smaller than
${1\over2}$. For large values of $ x $ we see that 
Eq.(\ref{menor1/4}) 
condition is satisfied. Therefore, (\ref{solfinal}) is a solution
when $ 0 < \alpha < {1\over4} $.

Note that when $ \alpha = {1\over 4} $ we have $\sigma = 0 $,
the constants $ a $ and $ b $ are identical, implying that
(\ref{assintotica}) cannot be used in this case. However, we
can perform the limit $ x \rightarrow \L^2$ already in the
differential equation (\ref{diferencial}) and study its behavior for this
particular value of $ \alpha $. In the asymptotic region (UV) we obtain
\begin{equation}
f(x)_{UV} = {x}^{-1/2}(C + D
\ln x). 
\label{igual1/4}
\end{equation}
We easily see that the ultraviolet boundary condition is satisfied for
large $ x $ and (\ref{solfinal}) is also a solution 
when $ \alpha = {1\over 4} $.
	 	
When $ \alpha >{1\over 4} $ in the parameters $a$ and $b$ of 
Eq.(\ref{solfinal}) we make the substitution $\sigma \rightarrow i\rho $
with $\rho $ defined as $(\alpha - {1\over4})^{1\over2} $. The asymptotic
behavior in this case is obtained with the help of (\ref{zero}) and 
(\ref{assintotica}) leading to 
\br
[(x&-&\a\kappa)f(x)]'|_{x\rightarrow \L^2} =  \nonumber\\ 
&-& 2  (\a \kappa )^{3/2}Re \bigg[ {\Gamma(2i\rho)\G(3/2 -\s)\G(3/2 +\s)\over
\Gamma^2({1\over2}+i\rho)}\bigg]\nonumber \\ 
&& \times \bigg({x\over\alpha \kappa } -1 \bigg)^{-{1\over
2}+ i\rho }\bigg|_{x\rightarrow \L^2}, 
\label{zeros}
\er
which is valid for $ x \gg \alpha \kappa $. Eq.(\ref{zeros})
has an infinite set of zeros located at
$ x = \alpha \kappa ( x_n + 1)$ for integer $n$, which can be
determined with the procedure of the second paper of Ref.~\cite{atkinson}.

The real part of Eq.(\ref{zeros}) can be written as
\begin{equation}
\exp \bigg\{\ln  \bigg[ {\Gamma(2i\rho)\G(3/2 -\s)\G(3/2 +\s)\over
\Gamma^2({1\over2}+i\rho)} x_n^{-{1\over
2}+ i\rho }\bigg] \bigg\},
\label{exp}
\end{equation}
as $ \ln z = \ln |z| + i ( \arg z \pm 2n\pi) $, where $ z $ is a
complex number and $ n = 0,1,2, \ldots $, the zeros of this function
will occur for
\begin{equation}
\ln (x_n ) \sim {1\over\rho} \bigg[ 2n\pi 
 - \arg \bigg( {\Gamma(2i\rho)\G(3/2 -\s)\G(3/2 +\s)\over
\Gamma^2({1\over2}+i\rho)}\bigg) \bigg]
\label{log}
\end{equation}
with $ n= 1,2,\ldots $. This relation was obtained for
$ x \gg \alpha \kappa $, and $ n = 0 $ was excluded in
order to obtain only positive values in Eq.(\ref{log}). Therefore,
also for $ \alpha >{1\over 4} $ we do have a nontrivial solution.

In summary, when the gluon propagator is modeled by a delta function plus
a ``perturbative'' $1/k^2$ propagator we verified that the
chiral symmetry is broken for any positive value of the coupling
constant. One may ask if this result changes if we modify the
non-perturbative as well as the perturbative propagator, as in the 
case of Eq.(\ref{e2}), or with the inclusion of the
running coupling or the effect of a dynamical gluon
mass~\cite{atkinson,natale}. We will show in the next section that this is
not the case.

\section{An Infrared Model in the Gaussian Form}       

As discussed in the introduction a different ansatz for the
gluon propagator, exemplified by Eq.(\ref{e3}), has frequently been used.
Its non-perturbative infrared behavior is given by the first term of
Eq.(\ref{e3})
\begin {equation}
 D_{\mu\nu}^{ab}(k)= \delta^{ab}\bigg[ \delta_{\mu\nu} - {k_{\mu} k_{\nu} \over
{k^2} }\bigg]3\pi^2{\chi^2\over \Delta^2}\exp {-k^2\over\Delta}, 
\label{gauss}
\end{equation}
where the parameter $ \chi $ controls the intensity of the interaction,
and  $\Delta $ gives the gaussian width. It is important to note that
in the limit $ \Delta  \rightarrow 0 $  we recover the infrared 
behavior of the propagator discussed in the previous section.

In this section we will show that the quark self-energy calculated
with the propagator of Eq.(\ref{gauss}) has a well defined bifurcation
point. Afterwards, we verify that adding a perturbative tail to this
propagator results in a smaller critical coupling constant than the one
obtained only with the perturbative part. Finally, in the limit $ \Delta 
\rightarrow 0 $, we recover the result of the previous section showing that it
is independent of the perturbative part that is added to the delta function.

To find nontrivial small solutions of the nonlinear Schwinger-Dyson
equation for the quark self-energy we can study the linearized 
equation. The substitution of Eq.(\ref{gauss}) into 
Eq.(\ref{edse}) gives the following linear integral equation for
$f(x)$ (recalling that $f(x)= \delta B(x)$)
\begin{equation}
f(x)= {3\over 4} g^2 {\chi^2\over\Delta^2}\bigg[\int^x_0 dy
f(y)\exp{-x\over\Delta} + \int_x^{\infty} dy f(y)\exp{-y\over\Delta}\bigg],
\label{gauss2}
\end{equation}
where we considered $ A(p^2) = 1 $, since, according to bifurcation theory,
the equation for $(A(p^2) - 1)$ is of higher order in the functional
derivative of $B(p^2)$, and the critical coupling constant is
determined only through Eq.(\ref{gauss2}). To obtain Eq.(\ref{gauss2}) we
performed in the gluon propagator the so called angle approximation, which is
given by   
\begin{equation}
D( (p-q)^2) \approx \theta(p^2 -q^2)D(p^2) + \theta(q^2-p^2)D(q^2). 
\label{angle-approx}
\end{equation}

The Eq.(\ref{gauss2}) is a homogeneous
Fredholm equation with the kernel %
\begin{equation}
K_I(x,y) = \exp{\left(-{x\over\Delta}\right)} \theta(x-y) +
\exp{\left(-{y\over\Delta}\right)} \theta(y-x). 
\label{kernel}
\end{equation}
The norm of (\ref{kernel}) is easily calculated
\begin{equation}
\parallel K_I \parallel ^2 = \int_0 ^\infty dx\int_0 ^\infty dy K^2_I(x,y) =
{\Delta^2\over 2},
\label{norma}
\end{equation}
and we find nontrivial $L^2$ solutions of $B(x)$ for $g^2$
on a point set whose smallest positive point satisfies~\cite{atkinson2}
\begin{equation}
g^2  \geq \frac{4 \Delta^2}{3 \chi^2} \,
\frac{1}{\parallel K_I\parallel } = \left( \frac{4\sqrt{2}}{3} \right)
\frac{\Delta}{\chi^2} \approx 1.88 \, \frac{\Delta}{\chi^2}. 
\label{iq1}
\end{equation}
Eq.(\ref{iq1}) gives a lower bound for the critical
point $g^2_c$. However, we can do better than this.  Using the method of
traces we can show that the approximate critical
value is indeed of the order of the smaller value of the bound given by
Eq.(\ref{iq1}). 

The following approximate formula holds true for the smallest
characteristic number $g^2_c$~\cite{integrais} 

\begin{equation}
|g^2_c|  \approx  \sqrt{\frac{A_2}{A_4}},
\label{iq2}
\end{equation}
where for a symmetric kernel
\begin{equation}
A_{2m} = 2\int_0^{\infty}\int_0^{x} \, K^2_m (x,y) \, dy \, dx ,
\label{iq3}
\end{equation}
with $m$ running over $ 1,2 $. $K_2 (x,y)$ is given by
\begin{equation}
K_2 (x,y) = \int_0^{\infty} \, K_1 (x,z) K_1 (z,y) \, dz ,
\label{iq4}
\end{equation}
and, in the case of the infrared part of the propagator,
$K_1 (x,y) = (3\chi^2/4\Delta^2) g^2 K_I(x,y)$. The calculation of
Eq.(\ref{iq2}) entails 
\be
g^2_c \approx \sqrt{\frac{9\chi^4/32\Delta^2}{297\chi^8/4096\Delta^4}}
\approx 1.97 \frac{\Delta}{\chi^2}.
\label{iq5}
\ee
It is known that this method overestimates $g_c^2$. In this case
we verify that the smallest value of our lower bound Eq.(\ref{iq1})
is a good approximation for the critical point.
The value of the critical coupling can be obtained substituting the
phenomenological values of $\chi$ and $\Delta$ into Eq.(\ref{iq1}). 
It is obvious that in the limit $ \Delta  \rightarrow 0 $ we recover
the result of the previous section, {\sl i.e.} with the gluon propagator
given by a $\delta$ function the chiral symmetry is broken for
any coupling constant $g^2 > 0 $.
	
Let us now consider the case where we add to the gaussian form
of the infrared propagator a perturbative contribution. If this
new contribution is of the form given by (\ref{e2}) or (\ref{e3}), 
it is not difficult to verify the existence of a critical coupling 
constant for the onset of chiral symmetry  breaking. The reason for this is
that the ultraviolet behavior of the  propagator is softened by the 
factors $1-\exp (-k^2/[4m_t^2]) $ and  $ {\cal{F}}(k^2)$, producing an effect
equivalent to a dynamical gluon mass~\cite{silva} for which it was shown the
existence of a bifurcation point~\cite{atkinson,natale}. Therefore, we proceed
to the most intriguing case where we add to the ansatz for the
infrared gluon propagator the contribution of QCD with massive gluons
and the effect of the running coupling constant,
{\sl i.e.} a propagator proportional to  
$\alpha_s(k^2)/(k^2+m_g^2) $. 

If we denominate by $ K_U$ the kernel related to this
ultraviolet propagator, we assume that we do have
a nontrivial solution for
\begin{equation}
g^2 \geq  \frac{1}{\parallel K_U \parallel}  .
\label{gq3}
\end{equation}
A lower bound for the bifurcation point in the case that we consider the sum of
propagators will be given by 
\begin{equation}
{g^2_c}\parallel \frac{3 \chi^2}{4 \Delta^2}   K_I + K_U \parallel \geq 1,
\label{cond}
\end{equation}
where $ \parallel K_I \parallel $ is given
by Eq.(\ref{norma}).
With the triangle inequality
\begin{equation}
\parallel K_A + K_B \parallel \le \parallel K_A \parallel + \parallel K_B
\parallel,      
\label{destriang}
\end{equation}
and making use of Eq.(\ref{norma}) we obtain
\begin{equation}
{g^2}\bigg({3\sqrt{2}\over 8}{\chi^2\over \Delta}
+\parallel K_U\parallel \bigg)\geq 1.
\label{condfin}
\end{equation}

The solutions that break the chiral symmetry appear
only for values of the coupling constant obeying Eq.(\ref{condfin}). Note
that as long as we can factor out $g^2$ from both propagators, and as long
as the kernel of the perturbative tail is bounded, we obtain a condition for
values of $g^2$ above which the self-energy bifurcates. The lower bound
for the critical value
\begin{equation}
{g^2_c} \geq \frac{1}{\bigg({3\sqrt{2}\over 8}{\chi^2\over \Delta}
+\parallel K_U\parallel \bigg)} ,
\label{cfin}
\end{equation}
approaches zero as we
take the limit $ \Delta  \rightarrow 0 $, 
and is compatible with the result of the previous section. 

Considering the infrared and ultraviolet contributions to
the propagator, we apply again the method of traces to determine 
the critical coupling constant through Eqs.(\ref{iq2}) and (\ref{iq3}).
In this case the kernel $K_1 (x,y)$ in Eq.(\ref{iq3}) is given by
\be
K_1 (x,y) = H(x) \theta (x-y) + H(y) \theta (y-x),
\label{kiu}
\ee
where
\be
H(x) = \frac{3 \chi^2}{4 \Delta^2} \exp \left(\frac{-x}{\D}\right) + 
\frac{1}{x + m_g^2} 
\frac{4 \pi^2 d}{ \ln {(\frac{x}{\Lambda^2} + \tau)}}.
\label{hx}
\ee
Note that we have already factored out from the kernel
the coupling $g^2$.
To determine $A_4$ we need $K_2 (x,y)$ which is equal to
\br
K_2 (x,y) = H(x)H(y) \int_0^y dz + H(x) \int_y^x dz \, H(z) \nonumber\\
 + \int_x^{\infty} dz \, H^2 (z).
\label{k2iu}
\er

The critical coupling constant was determined numerically for
four quark flavors, $\chi^2 = 2.0 \, \mbox{GeV}^2$, $m_g = 600 \,$ MeV, 
$\Lambda = 300 \,$ MeV, and we assumed $\tau = 16$, where all
these values are phenomenologically acceptable, and, in particular,
the value of $\tau$ is compatible with the infrared behavior of the 
running coupling constant in a theory admitting dynamically
generated gluon masses~\cite{cornwall} (in these papers
we have $\tau \approx 4 (m_g^2/\Lambda^2)$). The critical
coupling is shown in Fig.(1) as a function of the parameter
$\Delta$. Note that as we decrease  $\Delta$
the value of the critical coupling goes to zero, confirming
the result of the previous section. In this
region also the integrals appearing in Eq.(\ref{k2iu}) become
more problematic because the gaussian becomes very peaked. Changes in the form
of the ultraviolet propagator barely affect this result. Finally, we
confirm that for this class of infrared propagators the chiral phase
transition happens at a smaller value than the one obtained with perturbative QCD.

 \begin{figure}
 \setlength{\epsfxsize}{1.0\hsize} \centerline{\epsfbox{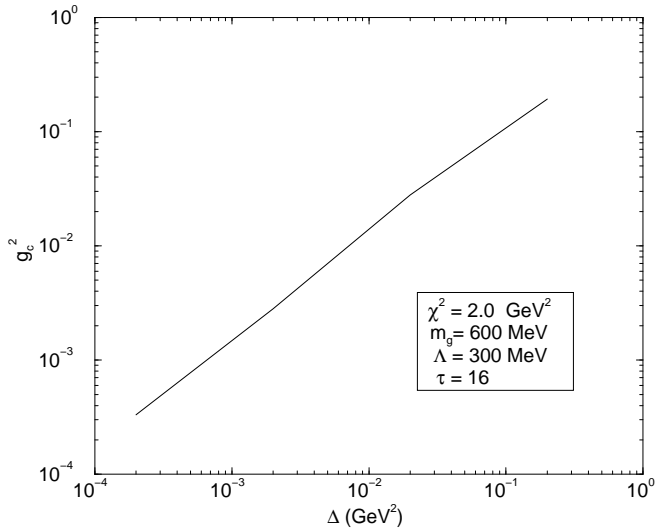}}
 \caption[dummy0]{Critical coupling constant, $ g_c^2 $,  as a function of
the parameter $\Delta$, considering the infrared and ultraviolet
contributions to the gluon propagator. We see that decreasing the value of
$\Delta$ the critical coupling goes to zero. The evaluation was perfomed for 
$n_f = 4$, $\chi^2 = 2.0 \, \mbox{GeV}^2$, $m_g = 600 \, \mbox{MeV}$,  
$\Lambda = 300 \, \mbox{MeV}$ and $\tau = 16$.} \end{figure}

\section{Conclusions}

An ansatz for
the infrared gluon propagator in the form of a $ \delta^4(q) $ function
(or a variation thereof), frequently used in applications of the
Global Color Model, gives an
excelent description of hadronic properties connected with chiral symmetry
breaking. In this work we studied the bifurcation of the quark self-energy
within this model. The idea was to compare GCM to QCD with just the
perturbative gluon propagator, where it is known that above a certain critical
value of the coupling constant the chiral symmetry is 
broken~\cite{atkinson,roberts,peskin}. We verified  that the introduction of a
delta function to describe the infrared behavior of the gluon propagator
implies that the chiral symmetry is always broken for any value of the
coupling constant. This is an interesting result if we remember that
this propagator is considered to describe confined quarks~\cite{munczek}.
However, this is certainly in contrast with what is known to happen
in QCD.

With a model for the gluon propagator inspired by the delta function but 
softer at the origin (in a gaussian form), we verified, using
the method of traces, that the critical
coupling for the onset of chiral symmetry breaking is lower than the one
expected for perturbative QCD, and recovered in a particular limit the result
obtained with the model involving a delta function for the infrared gluon
propagator. Therefore, the critical point for chiral symmetry breaking may
distinguish among different QCD motivated models, and in the cases we
have studied the critical coupling is always smaller than the one known for
QCD~\cite{roberts,peskin}.

\section*{Acknowledgments}  

We would like to thank E.V.Gorbar for valuable remarks.
This research was supported by the
Conselho Nacional de Desenvolvimento
Cient\'{\i}fico e Tecnol\'ogico (CNPq) (AAN,RR),
by Fundac\~ao de Amparo \`a Pesquisa do Estado de 
S\~ao Paulo (FAPESP) (ACA,AAN,RR) and by Programa de
Apoio a N\'ucleos de Excel\^encia (PRONEX).

\end{document}